%% file: 00_main.tex
\documentclass[sigconf]{acmart}
\AtBeginDocument{%
  \providecommand\BibTeX{{%
    \normalfont B\kern-0.5em{\scshape i\kern-0.25em b}\kern-0.8em\TeX}}}

\setcopyright{acmcopyright}
\copyrightyear{2018}
\acmYear{2018}
\acmDOI{XXXXXXX.XXXXXXX}

\acmConference[Conference acronym 'XX]{Make sure to enter the correct
  conference title from your rights confirmation emai}{June 03--05,
  2018}{Woodstock, NY}
\acmBooktitle{Woodstock '18: ACM Symposium on Neural Gaze Detection,
 June 03--05, 2018, Woodstock, NY} 
\acmPrice{15.00}
\acmISBN{978-1-4503-XXXX-X/18/06}

\usepackage{xspace}
\newcommand{\system}{CellSync\xspace} 
\newcommand{\systems}{CellSync's\xspace}

\newcommand{\smallset}{SnapGrid\xspace} 
\newcommand{\smallsets}{SnapGrids\xspace} 
 
\newcommand{\smallsetnospace}{SnapGrid}

\usepackage{xcolor}

\usepackage{framed}

\usepackage{tabularx}

\newcommand*{\img}[1]{%
    \raisebox{-.3\baselineskip}{%
        \includegraphics[
        height=\baselineskip,
        width=\baselineskip,
        keepaspectratio,
        ]{#1}%
    }%
}

\usepackage{listings}
\newenvironment{promptbox}{\begin{oframed}}{\end{oframed}}

\copyrightyear{2024}
\acmYear{2024}
\setcopyright{rightsretained}
\acmConference[CHI EA '24]{Extended Abstracts of the CHI Conference on Human Factors in Computing Systems}{May 11--16, 2024}{Honolulu, HI, USA}
\acmBooktitle{Extended Abstracts of the CHI Conference on Human Factors in Computing Systems (CHI EA '24), May 11--16, 2024, Honolulu, HI, USA}
\acmDOI{10.1145/3613905.3651115}
\acmISBN{979-8-4007-0331-7/24/05}

\begin{document}

\title[LLMs for Domain Expert Inclusion in Data Science]{Leveraging Large Language Models to Enhance Domain Expert Inclusion in Data Science Workflows}


\author{Jasmine Y. Shih}
 \affiliation{%
  \institution{Stanford University}
   \country{USA}
 }
 \email{jyshih@stanford.edu}
 
\author{Vishal Mohanty}
 \affiliation{%
  \institution{Stanford University}
   \country{USA}
 }
 \email{vmohanty@stanford.edu}

\author{Yannis Katsis}
 \affiliation{%
  \institution{IBM Research}
   \country{USA}
 }
 \email{yannis.katsis@ibm.com}
 
\author{Hariharan Subramonyam}
 \affiliation{%
  \institution{Stanford University}
   \country{USA}
 }
 \email{harihars@stanford.edu}

\renewcommand{\shortauthors}{Shih, et al.}

\begin{abstract}
Domain experts can play a crucial role in guiding data scientists to optimize machine learning models while ensuring contextual relevance for downstream use. However, in current workflows, such collaboration is challenging due to differing expertise, abstract documentation practices, and lack of access and visibility into low-level implementation artifacts. To address these challenges and enable domain expert participation, we introduce \system, a collaboration framework comprising (1) a Jupyter Notebook extension that continuously tracks changes to dataframes and model metrics and (2) a Large Language Model powered visualization dashboard that makes those changes interpretable to domain experts. Through \systems cell-level dataset visualization with code summaries, domain experts can interactively examine how individual data and modeling operations impact different data segments. The chat features enable data-centric conversations and targeted feedback to data scientists. Our preliminary evaluation shows that \system provides transparency and promotes critical discussions about the intents and implications of data operations.
\end{abstract}





\maketitle

\input{01_intro}
\input{02_relatedwork}
\input{04_ux}
\input{05_system}
\input{06_evaluation}
\input{08_discussion}
\begin{acks}
This work was partially supported by IBM as a founding member of Stanford Institute for Human-centered Artificial Intelligence (HAI).
\end{acks}

\bibliographystyle{ACM-Reference-Format}
\bibliography{99_refs}

\input{appendix}

\end{document}

%% file: 01_intro.tex
\section{Introduction}

When building machine learning (ML) models, collaborating with domain experts such as educators, public health experts, agricultural specialists, etc. is essential to ensure the accuracy, fairness, and contextual applicability of ML models~\cite{viaene2013data}. For instance, the domain expert might provide knowledge about specific data patterns useful for missing value imputations; based on current model performance, they may suggest creating new features such as calculating blood sugar levels over a 3-month period. Domain experts play a critical role in ensuring ML solutions are tailored to the broader domain goals, including fairness, transparency, and accountability~\cite{nourani2020role, whang2023datacentricai}.

Given the dynamic and iterative nature of ML modeling processes, cross-discipline collaboration should be \textit{continuous} through constant evaluation and feedback~\cite{jobin2019ethicalai, subramonyam2022solving}. However, current support for domain expert involvement is largely limited to abstract summarizations of modeling decisions in the form of documentation. Artifacts such as Datasheets~\cite{gebru2021datasheets}, Factsheets~\cite{arnold2019factsheets}, and Model Cards~\cite{mitchell2019model} are only created towards the end of the development process~\cite{heger2022}, making it difficult for domain experts to actively participate in upstream modeling tasks. Furthermore, domain experts not proficient in data-driven work may face challenges in accessing and interpreting computational artifacts such as code~\cite{wong2018de}. This knowledge gap necessitates extra work from engineering teams to translate complex decisions into formats accessible to domain experts. However, a lack of shared understanding of each other's tasks and terminology can lead to misunderstandings, inconsistent assumptions, and even conflicts~\cite{nahar2022collab}.

To address current challenges in collaborative ML, we explore the potential of large language models (LLMs) to enhance transparency and facilitate the involvement of domain experts in data science workflows. We developed \system, which consists of a visualization dashboard powered by LLMs that provides domain experts with an `alternate' view of computational notebooks commonly utilized by data science practitioners. Complementing the visualization interface is a Jupyter Notebook \cite{jupyter} extension that parses and relays computational operations to the dashboard. As the data scientist performs each data operation in Jupyter Notebook, a data version card containing a text summary and a grid-based visualization (\textit{\smallset}) is added to the dashboard to explain the data operation. \system leverages LLM one-shot, few-shot, and chain-of-thought prompting \cite{min2022fewshot, ahmed2023llm, wei2022chain} to generate text summaries of Python code and extract information for visualization purposes. Additionally, \system supports a chat feature between the dashboard and the notebook for domain experts to directly share feedback with data scientists. In the current instantiation of our work, we focused on tabular data given their widespread prevalence in predictive modeling tasks. To evaluate how \system supports domain experts' understanding of ML data operations, we conducted a controlled study with 10 pairs of domain experts and data scientists. Our work contributes 1) an LLM-powered visualization dashboard that makes it possible for domain experts to follow along with data scientists' work in Jupyter Notebook and 2) a preliminary evaluation of the dashboard features.

%% file: 02_relatedwork.tex
\section{Related Work}

In collaborative ML, researchers have investigated how data science and ML practitioners may work with domain experts in different stages of the ML pipeline ~\cite{passi2018collab, mao2019collab, wang2022collab, deng2023collab, nahar2022collab, piorkowski2021, sambasivan2021, moller2023}. Current collaboration approaches are largely through live meetings, messaging channels, and, in a few instances, specially designed graphical interfaces~\cite{piorkowski2021, deng2023collab, park2021, macinnes2010, du2010, farahani2019}. Collaborative ML teams engage in a mix of synchronous and asynchronous communication tools depending on whether the current discussion requires content-based (e.g. Jupyter Notebook and Git version control systems) or process-based communication (e.g. emails and Slack) \cite{roy2023dscommunication, mao2019collab}. To ensure effective collaboration, data scientists and domain experts must continue to establish content common ground (shared understanding of work subject) and process common ground (shared understanding of procedures) throughout the lifecycle of ML tasks ~\cite{mao2019collab}. However, the lack of centralized tools and protocols makes it difficult for multidisciplinary teams to sustain collaboration across ML stages. As a result, the interactions between data scientists and domain experts are often limited to one stage of the ML pipeline~\cite{sambasivan2021, moller2023}.

Differences in expertise between domain experts and data scientists pose additional challenges. Data scientists often spend large amounts of time preparing concrete data examples to communicate with domain experts~\cite{park2021}, and meaningful ML stakeholder participation requires scaffolds to aid the exchange of technical and non-technical considerations~\cite{tan2023educationml}. To bridge the expertise gap, systems should be designed to aid domain experts in understanding data science techniques in a manner that better matches their mental model of data~\cite{jung2022}. Visualizations of data changes~\cite{lucchesi2022smallset, niederer2018vizchange, hohman2020} have proved to be an effective means of supporting understanding beyond serving the function of documentation. Recent work in Explainable AI~\cite{doran2017, dwivedi2023, gade2019, holzinger2017, wang2022interpretability} has also demonstrated ways to make ML components more interpretable to non-experts through visualization and direct manipulation interactions~\cite{mohseni2021, alicioglu2022, pereira2020}.

Informed by the challenges and potential solutions identified in prior research, we explore the use of LLMs to supplement human effort in translating code into natural language and explainable visualizations. In \system, we build on the Smallset Timelines visualization technique~\cite{lucchesi2022smallset} to select and visualize a digestible subset of rows and columns for domain experts. Further, we utilize LLMs' code summarization capabilities~\cite{ahmed2023llm} and refine these summaries through targeted prompt engineering to tailor them towards domain experts. To better support data scientists in their collaboration with domain experts, we looked to prior works to design the data-scientist-facing component of \system. Inspired by the Callisto system~\cite{wang2020} that enables data scientists to work together in computational notebooks with context-rich chat features, as well as studies that reveal collaboration patterns among data scientists~\cite{wang2019, raghunandan2023}, we developed a Jupyter Notebook extension that is situated within data scientists' ML work environment and provides a chat interface for data-centric communication with domain experts. In our work, we focus specifically on tabular data due to its high availability in health and education domains. We selected predictive model development as the ML task to design for because predictive models have high implications in our chosen domains~\cite{gianfrancesco2018bias, obermeyer2019, huang2022,gardner2019bias, baker2021}.

%% file: 04_ux.tex
\section{User Experience}

\begin{figure*}[!htbp] 
  \centering
  \includegraphics[width=\linewidth]{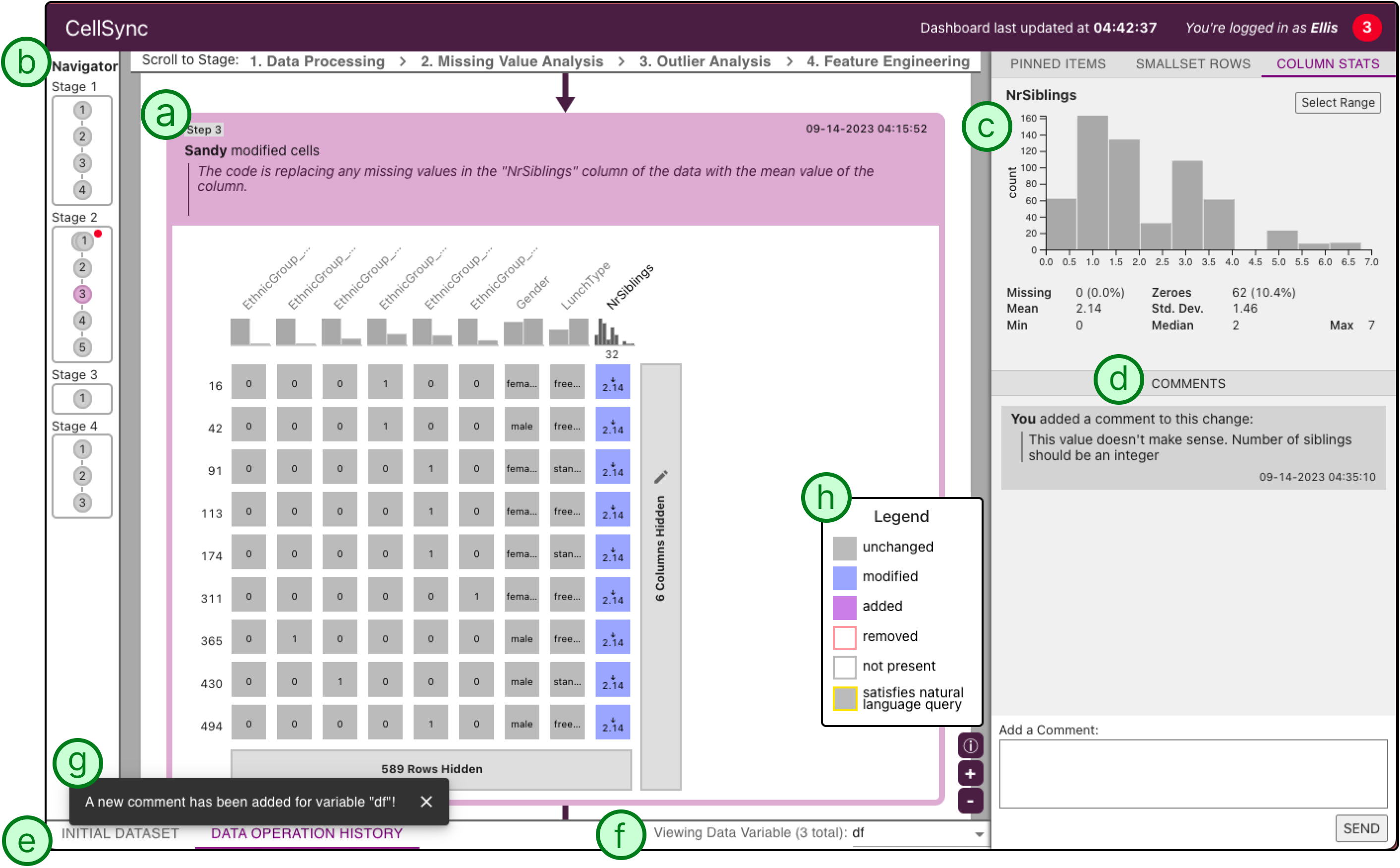}
  \caption{The \system visualization dashboard interface: (a) data version card containing code summary and \smallset along with column histograms, (b) clickable card navigator displaying a red dot indicating new comments, (c) detailed view for the selected column, (d) comments section for the current card, (e) bottom bar for switching between initial dataset table and data operation history, (f) dropdown menu for variable selection, (g) notification for new comments, and (h) collapsible legend for \smallset.}\label{figure:cellsync_viz_1}
  \Description{A screenshot of the visualization dashboard labeled with (a) to (h). (a) points to the data version card which contains a code summary at the top and a data subset visualization in the body. (b) points to navigator bar on the left side which represents each data version card in the history with a numbered circle. (c) points to the "Column Stats" tab in the right side panel displaying a histogram of a numeric column along with statistics such as mean, median, and standard deviation of the column. (d) points to the comment section in the right side panel where an existing comment is displayed and a text box is provided for sending a new comment. (e) points to the left portion of the bottom bar that provides the option to switch between "Initial Dataset" and "Data Operation History" views. (f) points to the right portion of the bottom bar that provides a dropdown menu for selecting a different variable to view. (g) points to a small pop-up message in the lower left corner of the dashboard with the text "A new comment has been added for variable 'df'!" (h) shows the legend for the data subset visualization, including gray for unchanged values, blue for modified values, purple for added values, red border for removed values, gray border for values not present, and yellow border for values that satisfy the natural language query.}
\end{figure*}

\begin{figure}
  \centering
  \includegraphics[width=\linewidth]{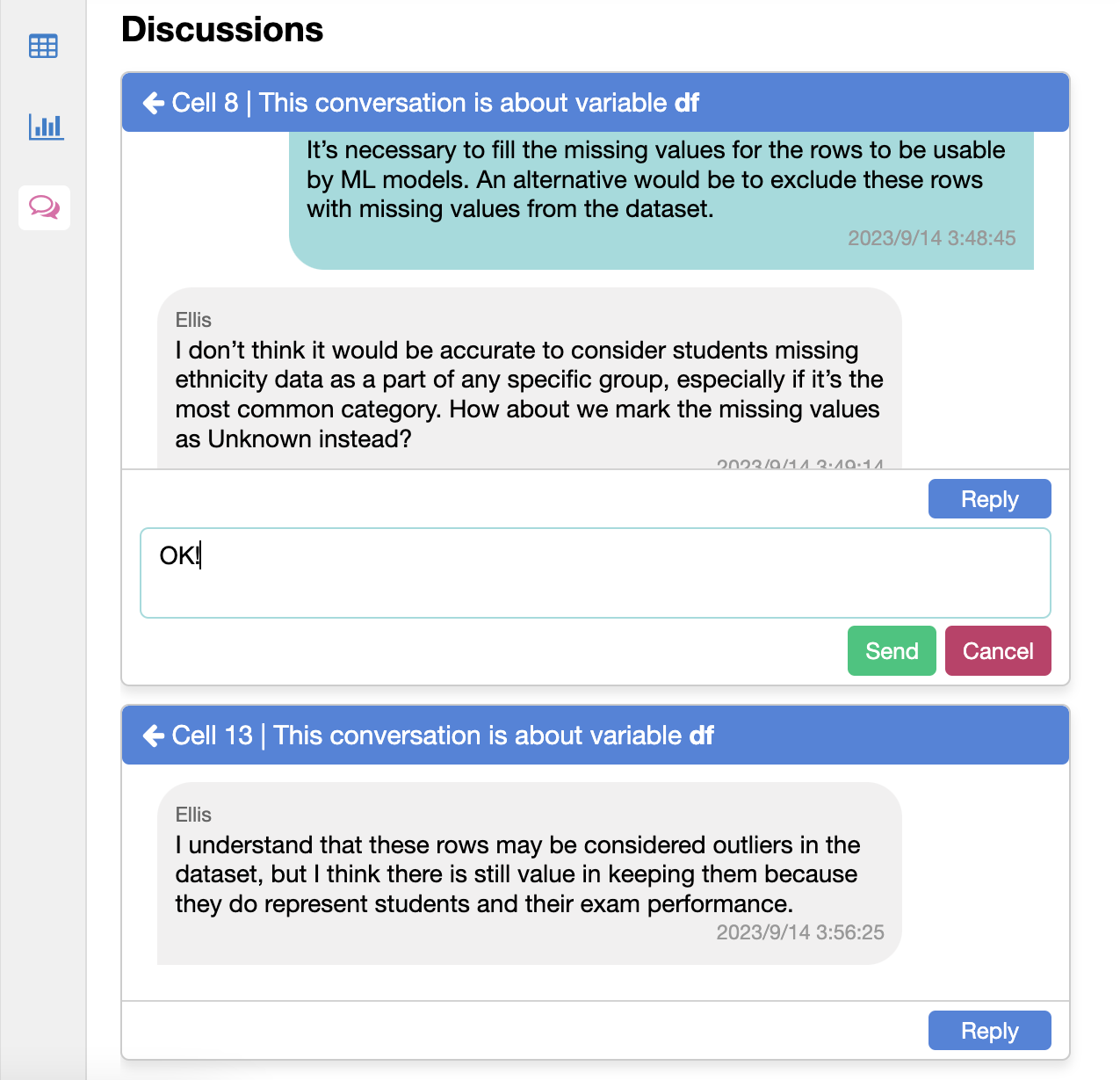}
  \caption{The \system data-scientist-facing chat interface rendered by the Jupyter Notebook extension.}\label{figure:jupyter_extension_chat}
  \Description{A screenshot of the side panel in Jupyter Notebook rendered by a Jupyter Notebook extension showing a vertical stack of three icons (table, chart, and dialogue, which is currently highlighted) on the left bar and a chat interface on the right side showing back-and-forth conversation threads between Sandy and Ellis regarding specific data operations.}
\end{figure}

\system consists of a Jupyter Notebook extension and a web visualization dashboard. The extension tracks data scientists' data operations in each notebook cell. The visualization dashboard, designed for domain experts, communicates each change to a dataframe with a text summary and a grid-based visualization (\textit{\smallset}) on a data version card (Figure~\ref{figure:cellsync_viz_1}a). Both the extension and the dashboard provide a chat feature for domain experts (Figure~\ref{figure:cellsync_viz_1}d) and data scientists (Figure~\ref{figure:jupyter_extension_chat}) to exchange comments. Each \smallset, inspired by Smallset Timelines~\cite{lucchesi2022smallset} which visualizes data changes in grid-based snapshots, shows a static 9-by-9 subset of the data; the subset is selected to maximize the coverage of the changes tied to the data operation. On each square \img{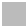} that represents a cell in the dataframe, the cell's previous and new values are displayed with an arrow between them. Cells, rows, or columns that have been modified, added, or removed are highlighted with different colors -- blue \img{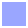} for modification of cells, purple \img{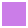} for new columns and boxes \img{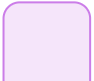} around related columns, and red \img{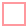} for removed columns or rows. A navigator on the left of the page provides a compact view of the data operation history with clickable circles for navigation (Figure~\ref{figure:cellsync_viz_1}b).

To illustrate \systems features and user experience, we present a fictional journey of Sandy, a data scientist, and Ellis, an education domain expert, who work together on an ML model to predict students' writing scores.
\vspace{1mm}

\noindent
\textbf{Set-up:} Sandy obtains a dataset with 600 rows of student background information and exam scores. She loads the dataset into a \textit{pandas dataframe} variable named \texttt{df} by executing a Jupyter Notebook cell. In the web dashboard, Ellis sees a text summary about the operation along with a \smallset showing a subset of the loaded data in the corresponding data version card. He clicks on the thumbnail histogram \img{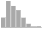} under each column name on the card to open the histogram and column statistics in the side panel (Figure~\ref{figure:cellsync_viz_1}c).
\vspace{1mm}

\noindent
\textbf{Missing Value Analysis:} Sandy fills in the missing values in the \textit{``EthnicGroup''} column, which contains categorical values, with the most frequent category. In the web dashboard, Ellis attempts to understand the newly executed operation by reading the code summary and viewing the \smallset. As more than 9 values have been modified, all 9 cells in the column are colored in blue \img{figures/glyphs/modified}, and a number above the first cell in the column indicates the total number of values changed. Concerned about the implications of assuming missing data, Ellis sends a comment to Sandy to suggest she replace the missing values with ``Unknown'' instead.
\vspace{1mm}

\noindent
\textbf{Feature Encoding:} Sandy performs one-hot encoding on a categorical column, thereby creating a new column for each category and removing the original column from the dataframe. In the corresponding \smallset, the cells in the new columns are colored in purple \img{figures/glyphs/added.png}, and the cells in the removed column are colored in white with a red border \img{figures/glyphs/removed.png}. In addition, a light purple box \img{figures/glyphs/box.png} is placed around the removed column to indicate that the new columns have been created based on values in this original column. Ellis views the data version card for the one-hot encoding operation and understands that the encoding operation transforms the original categorical values into binary values.
\vspace{1mm}

\begin{figure*}[!h] 
  \centering
  \includegraphics[width=0.95\textwidth]{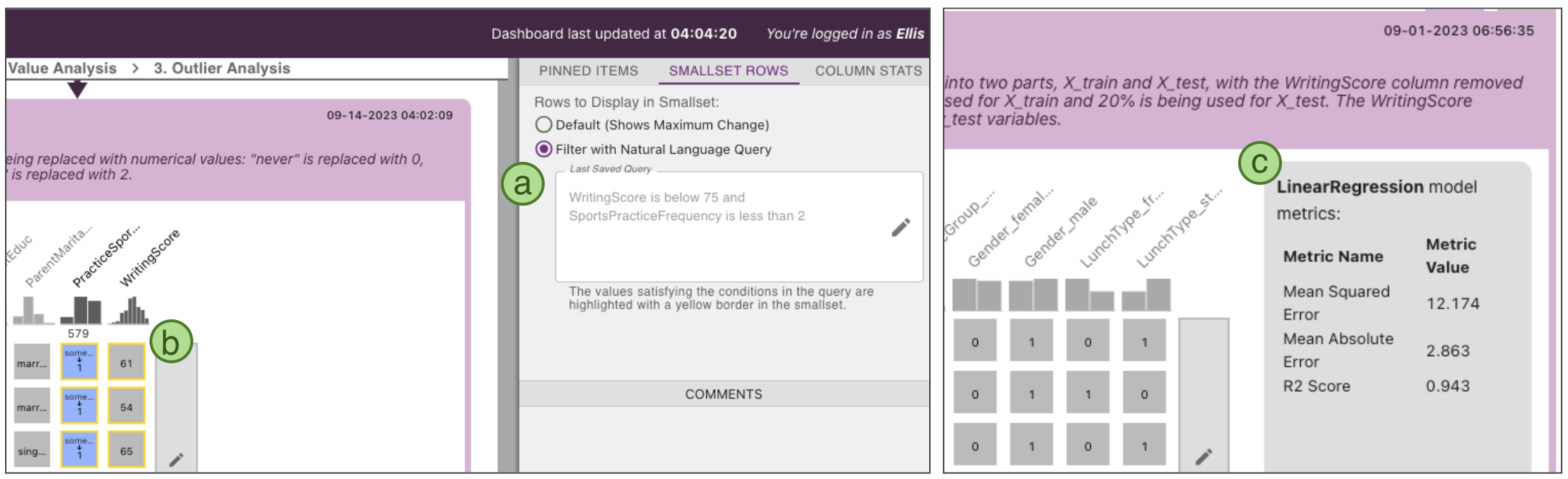}
  \caption{(a) Text field for entering a natural language query to update the selected \smallset. (b) Cells with values that meet the query criteria are highlighted. (c) Table display of model metrics information. }\label{figure:combined_model_metrics}
  \Description{Two screenshots of the system. The screenshot on the left shows a part of the side bar containing the text field for entering a natural language query to update the rows in the current SnapGrid. The screenshot on the right shows a part of a data version card containing a table that displays model performance metrics.}
\end{figure*}

\noindent
\textbf{Feature Selection:} Sandy observes that the \textit{``SportsPracticeFrequency''} column does not seem to be correlated with \textit{``WritingScore''} and decides to remove it. Ellis inquires Sandy about the deletion of data. Sandy explains that the column is not correlated with \textit{``WritingScore''} and therefore is not useful for the predictive model. To filter the raw data for his own examination, Ellis uses the natural language query feature in the side panel (Figure~\ref{figure:combined_model_metrics}a). He clicks on the previous data version card where the \textit{``SportsPracticeFrequency''} column is still present and submits the query \textit{``WritingScore is below 75 and SportsPracticeFrequency is less than 2''}. The updated \smallset highlights values that meet the criteria with a yellow border \img{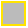} (Figure~\ref{figure:combined_model_metrics}b). This feature enables Ellis to examine the data in a flexible and natural manner.
\vspace{1mm}

\noindent
\textbf{Model Training:} Sandy splits the dataset into a training set stored as variable \texttt{X\_train} and a testing set stored as \texttt{X\_test}. The new dataframe variables lead to the addition of two new data version cards in the visualization dashboard. To access a new card, Ellis clicks on the variables drop-down menu in the bottom bar and selects the \texttt{X\_train} variable to switch the data operation history view to the new variable. Sandy then trains a \textit{``LinearRegression''} model with the \texttt{X\_train} variable and evaluates the model by calculating model performance metrics such as mean squared error, mean absolute error, and R2 score. In the visualization dashboard, Ellis is able to view the model name and metrics in the \texttt{X\_train} data version card (Figure~\ref{figure:combined_model_metrics}c).

%% file: 05_system.tex
\section{System Implementation}


\system consists of a Jupyter Notebook extension and web visualization dashboard. Both components of the system are synced with the shared Amazon DynamoDB database every 15 seconds using the serverless Amazon Lambda functions as the backend. The extension is also linked to  OpenAI's \emph{text-davinci-003} LLM model \cite{openai} to support several features in the backend.

\subsection{Jupyter Notebook Extension}
Our extension is built on the Jupyter Nbextensions framework\footnote{\url{https://github.com/ipython-contrib/jupyter\_contrib\_nbextensions}} and provides the core functionality to automatically share the data operation and model training history with the visualization dashboard. To capture the data operations in the ML model development process, we utilize a Python script that is invoked each time a code cell is executed in the notebook to obtain variable contents. Our script extends the script used in the open-sourced Variable Inspector extension\footnote{\url{https://jupyter-contrib-nbextensions.readthedocs.io/en/latest/nbextensions/varInspector/README.html\#variable-inspector}}, which lists all the variables in the context of the notebook using NamespaceMagics\footnote{\url{https://ipython.org/ipython-doc/2/api/generated/IPython.core.magics.namespace.html}}, to compare the current and prior states of dataframe variables. If a dataframe variable has been modified, we obtain an LLM-generated summary of the code (described in \ref{subsub: nl code description}) and compute a subset of the dataframe to be displayed in the visualization dashboard (described in \ref{subsub: subset_selection}). This information is packaged in a JSON object along with contextual data about the notebook cell. The JSON object is then sent to the Amazon DynamoDB database to allow information to be shared between the notebook and the visualization dashboard. In addition, each version of the full dataset is stored in our Amazon S3 database as a CSV file.

\subsubsection{Natural Language Code Description}\label{subsub: nl code description}
To automatically create text summaries of dataframe changes, we apply prompt engineering \cite{white2023prompt} to generate descriptions of the executed Python code. The prompt we engineered is given below, with \texttt{<dataframe\_var>} and \texttt{<code>} as placeholders for the variable name and code in the given notebook cell.\\

\begin{promptbox}
\begin{small}
\noindent
I want to explain a data transformation to a domain expert who may not be familiar with technical terms like `dataframe' or `transformation' as a very short summary. The goal is to make it easily understandable. Please explain what is happening to the data in the \texttt{<dataframe\_var>} variable. If any specific rows or columns are being modified, kindly mention them. It doesn't have to be too verbose. Avoid using terms like `dataframe'. If the code loads a dataset, do not infer what type of information is included in the dataset unless the information is exposed in the code.\\
\texttt{<code>}
\end{small}
\end{promptbox}

This prompt makes use of zero-shot prompting \cite{wei2022finetuned} with specific instructions aimed to make the LLM output more friendly to non-technical readers. In our prompt engineering process, we observed that the LLM tended to provide incorrect information about a dataset based solely on the dataset name present in the code (e.g. falsely stating what columns the dataset includes). To counter these LLM hallucinations \cite{zhang2023sirens}, we explicitly instruct the LLM to not infer the type of information included in a dataset referenced in the code.

\subsubsection{Computing Subset for \smallset}\label{subsub: subset_selection}
Building upon the Smallset Timelines subset computation \cite{lucchesi2022smallset}, we devise an algorithm to select a subset of 9 rows by 9 columns that maximizes the coverage of value changes for \smallset. The algorithm selects the dataset rows with the highest number of changed values, and the columns that have been directly affected or are implicitly involved in column changes (e.g. used to compute new columns). To determine the involved columns, we use the \emph{text-davinci-003} LLM to infer and extract the relationships between columns in the given code. In the LLM prompt, we first provide the executed code, followed by lists of existing columns and newly added columns in the dataframe, and then ask the LLM to return a JSON object indicating the relationships between the columns involved. We demonstrate the prompt with an illustrative example in the appendix (\ref{subsub:col_rel_prompt}). By leveraging the ability of the LLM to make inferences on column dependency using pattern recognition and data science knowledge, we obtain column relationships in a robust manner.

\subsubsection{Extracting Model Metrics}\label{subsub:extract_metrics}
When a model is trained on a dataframe variable and evaluated with performance metrics calculations, we extract the following model information using the \emph{text-davinci-003} LLM and the variable contents provided by the Python script: (1) Model Name: The name of the model, such as Linear Regression, (2) Train variables: The variables used in training, e.g. \texttt{X\_train, Y\_train}, (3) Test variables: The variables used for testing, e.g. \texttt{X\_test, Y\_test}, and (4) Metrics: A list of metric names (such as RMS error, Precision, Recall, Accuracy, etc.), their corresponding Python variable names, and the metric values. To get structured model metrics in the above format, we utilize few-shot chain-of-thought prompting \cite{wei2022chain} to \emph{teach} the LLM the kind of output we expect given the input. In the prompt, we first specify the desired output format and then provide several examples of input and expected output, followed by the code input that we need the LLM to generate output for. We found that, to increase the robustness of the LLM, it is necessary to provide a diverse set of code and output examples covering different models and performance metrics, as well as counterexamples that teach the model what types of input should result in empty output. The full prompt can be found in the appendix (\ref{subsub:model_metrics_prompt}).

\subsection{Visualization Dashboard}
The visualization dashboard is built with React.js for the frontend and serverless Amazon Lambda functions as the backend. The \smallset and the interactive column histograms are developed with the D3 library. We provide the domain expert with a natural and flexible mechanism to update the rows in the \smallset by submitting a natural language query. To achieve the desired outcome, we again apply few-shot chain-of-thought prompting to train the \emph{text-davinci-003} LLM model to transform the given natural language query to filter conditions (see full prompt in \ref{subsub:smallset_update} in the appendix). The LLM provides room for error in the natural language query as it understands the \emph{intention} of the domain expert in the context of the dataset. Even if the domain expert fails to make exact references to column names in their queries (incorrect case, added spacing, etc.), the LLM still correctly transforms the query to conditions that can be used to filter the rows in the backend.

%% file: 06_evaluation.tex
\section{Preliminary Evaluation}
In evaluating our system, our goal was to understand whether \system's features allow domain experts to understand the underlying data science code, and whether that understanding leads to communication and feedback to data scientists. We conducted a virtual user study on Zoom with pairs of domain experts and data scientists (a total of ten pairs). Five of the domain experts had an education background and the other five had public health expertise (see Table~\ref{tab:evaluation-study} in the appendix for participants' demographic information). The participants were recruited through our professional network and specialized communities on social platforms. All participants had at least 1.5 years of experience in their respective fields and received a \$60 gift card for two hours of their time.

In each session, we introduced the domain expert and the data scientist to a dataset relevant to the domain expert's field and an ML prediction task for them to collaborate on (see Table~\ref{tab:evaluation-study-descriptions} in the appendix for study details). To maximize the domain experts' use of the dashboard features, we placed the domain expert and the data scientist in two different Zoom rooms. After the main task, we brought the participants back to the same room and concluded the session with a 15-minute semi-structured interview. During the session, both participants' screens and the interview were recorded.

\subsection{Findings}
Overall, the domain expert participants found the code summary (see Table~\ref{tab:llm_code_summaries} in the appendix for examples) and the visualization features in the \system dashboard to be effective in aiding their understanding of data operations. Some participants preferred to read the text summaries to gain a high-level understanding of the data operation before deciding whether they needed to closely examine the \smallsetnospace. Other participants relied on the \smallset to understand the type and extent of data change. As the domain expert E7 commented, \smallsets helped her understand why it may be necessary for data scientists to perform certain operations:  \textit{``I really liked seeing the differences in the data highlighted\ldots to someone who isn't a data scientist, seeing the changes visualized this way helps bridge that gap of why an operation is important for a data scientist to do to make the data easier to work with.''} The participants’ differing preferences for the code summary and the \smallset suggest that displaying both components can support a wide range of domain experts in understanding data operations.

Most participants expressed that the visualization dashboard had a high learning curve. This sentiment stemmed from both the shortness of the evaluation session and the large number of features that the dashboard supported. Nonetheless, participants expressed appreciation for several features besides the code summary and the \smallsetnospace. In particular, the column histograms were extensively used. According to E3, \textit{``I can use them to make recommendations to the data scientist on what to pay attention to instead of relying on him to provide these basic statistics.''} In addition, domain experts found the data version card navigator to be helpful, as it clearly indicated the position of the current card and eliminated the need to scroll excessively in the history view.

Through \system, the participant pairs actively exchanged a variety of messages as they collaborated on the prediction task. For instance, many domain expert participants inquired about the purpose of one-hot encoding operations, and their collaborators would explain that it was a data science technique. The domain experts also raised concerns around missing value imputation, providing domain-specific reasons for why it may be
problematic. For example, public health experts argued that patients’ missing \textit{``Region''} data should not be imputed given the healthcare disparities between U.S. regions. In these cases, participant pairs engaged in discussions about the trade-offs between keeping the data true and making it usable by ML models, as they attempted to reach common ground. Our LLM-supported features helped domain experts to understand \textit{what changed} in the data, enabling them to initiate conversations with data scientists about \textit{why} certain data operations needed to be performed.

%% file: 08_discussion.tex
\section{Discussion and Conclusion}
To bridge the gap in expertise between domain experts and data scientists on ML collaboration tasks, we leveraged visualization and code summaries to help domain experts understand the impact of data science operations in the ML modeling pipeline. Further, the use of LLMs served multiple purposes in our system. First, the LLM-generated code summary reduces the communication burden on data scientists in the collaboration process, allowing them to spend more of their effort on data-specific communication rather than code-specific communication. Second, by using an LLM to extract dataframe- and model-related information from code, we utilize its capacity to understand the context and make inferences, increasing the system's robustness against a wide range of input.

Based on the preliminary evaluation findings, we aim to investigate how to apply different LLM tuning and prompting techniques to better support domain experts in our future work. One direction may be to contextualize LLM models in the given application domain, providing the LLMs with information about the collaboration task, objectives, and input data. As our education domain expert participant E1 explained that education experts typically approach education data with a standard set of considerations (e.g. correlation between socioeconomic factors and student performance), LLMs can be pre-instructed with domain-specific information to guide domain experts to more efficiently and thoroughly examine data transformations and model outputs. In addition, as research on using LLMs for code generation~\cite{li2023codeie, ouyang2023llm} begins to mature, we aim to increase our system's support for data scientists by providing automatic code suggestions based on domain experts' natural language feedback.

Our preliminary evaluation study was conducted over short periods of synchronous collaboration between the participants, resulting in limited iteration in the modeling process and may not reflect most collaboration and communication scenarios in reality. To evaluate our system in more realistic settings, we plan to deploy \system to industry teams that perform collaborative ML work and collect usage data over longer periods of time in our future research efforts. In addition, because \system can serve as a learning tool for data science education, we plan to integrate \system into university data science courses to examine its educational utility.

In conclusion, our system presents an approach to making computational notebooks accessible to domain experts, which can enhance their participation in upstream ML tasks. The use of LLMs for code summarization and data change visualizations can alleviate the engineering workload required to clarify decision-making processes and reduce friction in collaborative ML practices.

%% file: appendix.tex
\appendix

\section{Appendix}

\subsection{LLM Prompts}

\subsubsection{Column Relationship Extraction}\label{subsub:col_rel_prompt}

Below is an illustrative example of the prompt filled with sample code and lists of existing and newly added columns (in parts that follow a \#):\\
\begin{promptbox}
\noindent
\#Code
\begin{lstlisting}
df = pd.get_dummies(df, columns=["Gender"])
\end{lstlisting}
\#List of Existing Columns = \texttt{["Gender", "Age"]}\\
\#List of Newly Added Columns = \texttt{["Gender\_Female", "Gender\_Male"]}\\ \\
Are there new column(s) computed based on the existing columns in the code? If yes, provide the answer in the following format: {<new\_column>:[<existing columns used to compute this specific new column>]}. Remember to put double quotes (NOT single quotes) around the column names in the output format to make it a valid JSON string. If there are multiple detections, append the dictionary. If no, then just return an empty dictionary.\\
It is possible that new columns were generated by one-hot encoding techniques. In that case, make your best guess about which existing column was used to generate the new ones.\\
Answer:
\end{promptbox}

In this example, the LLM correctly returns \texttt{[\{"Gender\_Female": ["Gender"]\}, \{"Gender\_Male": ["Gender"]\}]}, even though the provided code does not explicitly show that the two new columns were computed based on the \textit{``Gender''} column.

\subsubsection{Model Metrics Extraction}\label{subsub:model_metrics_prompt}

The following is the full prompt we give to the LLM, with \texttt{<input\_code>} as placeholder for the code in the notebook cell we need the LLM to extract model information from.\newline

\begin{promptbox}
\noindent
I'll give you a piece of code used in an ML model, and you need to identify some metadata about the model in JSON format. The metadata includes `Model Name` (a string), `Train Variables` (a list), `Test Variables` (a list), `Metrics` (a list of objects each consisting of keys `Metric` and `Metric Variable` whose values are strings). For example, given\\
Input:
\begin{lstlisting}
# imports
import numpy as np
from sklearn.linear_model import LinearRegression
from sklearn.metrics import mean_squared_error
from sklearn.metrics import mean_absolute_error
# Generate some sample data
X_train = np.array([[1], [2], [3], [4], [5]])
Y_train = np.array([3, 5, 7, 9, 11])
# Train a linear regression model
reg =  LinearRegression().fit(X_train, Y_train)
# Test variables
X_test = np.array([6], [7])
y_test = ([13, 15])
y_test_pred = reg.predict(X_test)
# Calculate the mean squared error for the test data
mse_test = mean_squared_error(y_test, y_test_pred)
print("Mean squared error for test data:", mse_test)
# Calculate the mean absolute error for the test data
mae_test = mean_squared_error(y_test, y_test_pred)
print("Mean absolute error for test data:", mse_test)
\end{lstlisting}
Output:
\begin{lstlisting}
{
  "Model Name": "LinearRegression",
  "Train variables": ["X_train", "Y_train"],
  "Test variables": ["X_test", "y_test"],
  "Metrics": [
    {"Metric": "Mean Squared Error", "Metric Variable": "mse_test"},
    {"Metric": "Mean Absolute Error", "Metric Variable": "mae_test"}
  ]
}
\end{lstlisting}
\vspace{4mm}
Sometimes, the test variables might not be present. In that case, you can return an empty `Test variables`. For example given\\
Input:
\begin{lstlisting}
import numpy as np
from sklearn.linear_model import LinearRegression
from sklearn.metrics import mean_squared_error
# Generate some sample data
X_train1 = np.array([[1], [2], [3], [4], [5]])
Y_train1 = np.array([3, 5, 7, 9, 11])
# Train a linear regression model
reg1 = LinearRegression().fit(X_train1, Y_train1)
# Make predictions for the testing data
y_pred1 = reg1.predict(X_train1)
# Calculate the mean squared error
mse1 = mean_squared_error(Y_train1, y_pred1)
\end{lstlisting}
Output:
\begin{lstlisting}
{
  "Model Name": "LinearRegression",
  "Train Variables": ["X_train1", "Y_train1"],
  "Test Variables": [],
  "Metrics": [
    {"Metric":"Mean Squared Error", "Metric Variable": "mse1"}
  ]
}
\end{lstlisting}
\vspace{4mm}
Now, I'll give you another piece of code. Identify the metadata in it.\\
Input:
\begin{lstlisting}
from sklearn.linear_model import LogisticRegression
from sklearn.metrics import accuracy_score
from sklearn.metrics import precision_score
from sklearn.metrics import recall_score
from sklearn.model_selection import train_test_split
from sklearn.datasets import load_iris
# Load the iris dataset
iris = load_iris()
# Split the data into training and test sets
X_train, X_test, y_train, y_test = train_test_split(iris.data, iris.target, test_size=0.2, random_state=42)
# Train a logistic regression classifier
clf = LogisticRegression(random_state=42).fit(X_train, y_train)
# Make predictions for the test set
y_pred = clf.predict(X_test)
# Calculate the accuracy of the classifier
accuracy = accuracy_score(y_test, y_pred)
print("Accuracy:", accuracy)
# Calculate the precision of the classifier
precision = precision_score(y_test, y_pred)
print("Precision:", precision)
# Calculate the recall of the classifier
recall = recall_score(y_test, y_pred)
print("Recall:", recall)
\end{lstlisting}
Output:
\begin{lstlisting}
{
  "Model Name": "LogisticRegression",
  "Train Variables": ["X_train", "y_train"], "Test Variables": ["X_test", "y_test"],
  "Metrics": [
    {"Metric": "Accuracy", "Metric Variable": "accuracy"}, 
    {"Metric": "Precision", "Metric Variable": "precision"}, 
    {"Metric": "Recall", "Metric Variable": "recall"}
  ]
}
\end{lstlisting}
\vspace{4mm}
Very good! Here's another piece of code. Identify the metadata in it.\\
Input:
\begin{lstlisting}
import tensorflow as tf
from tensorflow import keras
import numpy as np
# Load the CIFAR-10 dataset
(x_train, y_train), (x_test, y_test) = keras.datasets.cifar10.load_data()
# Select a subset of the data
n_classes = 3
class_names = ['airplane', 'automobile', 'bird']
idx_train = np.isin(y_train, range(n_classes))
idx_test = np.isin(y_test, range(n_classes))
x_train, y_train = x_train[idx_train], y_train[idx_train]
x_test, y_test = x_test[idx_test], y_test[idx_test]
# Preprocess the data\nx_train = x_train / 255.0
x_test = x_test / 255.0
# Define the model architecture
model = keras.Sequential([
  keras.layers.Conv2D(32, (3, 3), activation='relu', input_shape=x_train.shape[1:]),
  keras.layers.MaxPooling2D((2, 2)),
  keras.layers.Conv2D(64, (3, 3), activation='relu'),
  keras.layers.MaxPooling2D((2, 2)),
  keras.layers.Flatten(),
  keras.layers.Dense(64, activation='relu'),
  keras.layers.Dense(n_classes)
])
# Compile the model
model.compile(optimizer='adam',
  loss=tf.keras.losses.SparseCategoricalCrossentropy(from_logits=True),
  metrics=['accuracy'])
# Train the model
model.fit(x_train, y_train, epochs=10, validation_data=(x_test, y_test))
# Evaluate the model on the test set
test_loss, test_acc = model.evaluate(x_test, y_test)
print('Test accuracy:', test_acc)
\end{lstlisting}
Output:
\begin{lstlisting}
{
  "Model Name": "Keras Sequential",
  "Train Variables": ["x_train", "y_train"],
  "Test Variables": ["x_test", "y_test"],
  "Metrics": [{"Metric": "Accuracy", "Metric Variable": "test_acc"}]
}
\end{lstlisting}
\vspace{4mm}
Note that you should only extract model meta data when an ML model is present. Do not treat methods in sklearn preprocessing modules used for scaling, normalization, or binarization, such as StandardScaler and RobustScaler (just to name a few), as ML models. Also do not treat transformers for missing value imputation, such as SimpleImputer and MissingIndicator as ML models. If no ML model is present in the code, simply return an empty object. For example, given\\
Input:
\begin{lstlisting}
from sklearn.impute import SimpleImputer
imp_mf = SimpleImputer(strategy='most_frequent', missing_values=np.nan)
for col in scores.drop(['Gender', 'MathScore'], axis=1).columns:
  scores[col] = imp_mf.fit_transform(scores[[col]])
\end{lstlisting}
Output:
\begin{lstlisting}
{}
\end{lstlisting}
\vspace{2mm}
- - - - -\\ \\
Now, I'll give you another piece of code. Identify the metadata in it.\\
Input:
\begin{lstlisting}
<input_code>
\end{lstlisting}
Output:
\end{promptbox}

The part before the short horizontal dashed line containing multiple pairs of input and output examples \emph{train} the LLM to return certain outputs, given certain inputs. By training the LLM through multiple steps, we are able to cover different types of inputs and corner cases before we ask it to extract model metrics on new input. The part after the dashed line contains the actual notebook cell code as input and the cue for the LLM to return the output.

\subsubsection{Transforming Natural Language Query to Filter Conditions}\label{subsub:smallset_update}

Similar to extracting model metrics in \ref{subsub:model_metrics_prompt}, we show the \emph{fine-tuning} part before the short horizontal dashed line and the \emph{evaluation} part after the dashed line in the prompt to the LLM, with placeholders for the column list and the natural language query denoted using \texttt{<columns>} and \texttt{<natural\_language\_query>}.\\

\begin{promptbox}
\noindent
I have a JSON object in Javascript, which represents a python dataframe. We need to get filtering conditions from it based on a input text. I'm going to teach you how to return a JSON object which is an array, each element containing `(columnName, condition, value)`. For example:\\
Columns: \texttt{[Glucose, Age, Gender, Outcome]} \\
Input: ``Show me rows/patients having glucose value > 90 and between the age of 25 to 35''\\
Output:
\begin{lstlisting}
[
  {"column": "Glucose", "operator": ">", "value": "90"},
  {"column": "Age", "operator": ">=", "value": "25"},
  {"column": "Age", "operator": "<=", "value": "35"}
]
\end{lstlisting}
\vspace{4mm}
Sometimes the `value` might be a string as well. For example:\\
Columns: \texttt{[Melatonin, Sickness level, Gender, Race, Predicted age]}\\
Input: ``Show me Female subjects whose melatonin is greater than 3.5''\\
Output:
\begin{lstlisting}
[
  {"column": "Gender", "operator": "==", "value": "Female"},
  {"column": "Melatonin", "operator": ">", "value": "3.5"}
]
\end{lstlisting}
\vspace{4mm}
Now lets do some practice, give me the output for:\\
Columns: \texttt{[Students Name, Education level, Parents education level, Dropped out]}\\
Input: ``Show me students whose parents' education level is High School and whose Dropped out is 1''\\
Output:
\begin{lstlisting}
[
  {"column": "Parents education level", "operator": "=", "value": "High School"},
  {"column": "Dropped out", "operator": "=", "value": "1"}
]
\end{lstlisting}

\vspace{2mm}
- - - - -\\ \\
Nice, now give me the output for:\\
Columns: \texttt{<columns>}\\
Input: \texttt{<natural\_language\_query>}\\
Output:
\end{promptbox}

\subsection{Evaluation Study Tables}\label{sub:tables}

\begin{table*}[h]
  \caption{Evaluation Study Participants Demographics}
  \label{tab:evaluation-study}
  \begin{tabular}{ccccccc}
    \toprule
    \multicolumn{4}{c}{Domain Experts} & \multicolumn{3}{c}{Data Scientists} \\
    \cmidrule(lr){1-4}\cmidrule(lr){5-7}
      & Domain & Gender & Experience (Years) & & Gender & Experience (Years) \\
    \midrule
    E1 & Education & F & 10 & S1 & M & 2 \\ 
    E2 & Education & F & 15 & S2 & M & 4 \\
    E3 & Education & F & 5 & S3 & M & 3 \\
    E4 & Education & F & 15 & S4 & M & 10 \\
    E5 & Education & M & 10 & S5 & M & 2 \\
    E6 & Public Health & F & 5 & S6 & M & 1.5 \\
    E7 & Public Health & F & 5 & S7 & M & 2 \\
    E8 & Public Health & F & 3 & S8 & M & 5 \\
    E9 & Public Health & F & 2 & S9 & M & 5 \\
    E10 & Public Health & F & 4 & S10 & M & 4 \\
    \bottomrule
\end{tabular}
\end{table*}

\begin{table*}[h]
    \centering
    \caption{Descriptions of Evaluation Study Prediction Tasks and Datasets By Domain}
    \label{tab:evaluation-study-descriptions}
    \small
\begin{tabular}{ m{0.2\textwidth} m{0.35\textwidth} m{0.35\textwidth}  }
\toprule
 & \textbf{Public Health Domain} & \textbf{Education Domain} \\ 
\midrule
Prediction Task  & Predict whether hospitalized patients will pass away due to adverse covid vaccine reaction  & Predict students' writing exam scores \\
\hline
Dataset Source \& Modification & Vaccine Adverse Event Reporting System (VAERS) dataset \cite{garg2021vaccinedata}, filtered down to entries corresponding to covid vaccine reactions and hospitalized patients for the year 2021, followed by random sampling for subset extraction & Fictional high school student dataset incorporating missing values and realistic socio-economic factors \cite{des2023studentdata}, randomly sampled to extract a subset \\
\hline
Dataset Dimension & 600 rows and 12 columns  &  600 rows and 12 columns \\
\hline
Feature Columns & Patient background information such as patient age, U.S. region, sex, disability, concurrent medications, etc.; Vaccine reaction data such as days of hospitalization, number of days to onset, vaccine manufacturer, ER visit, etc. & Student background information such as gender, ethnic group (as category A to E), parent's education level, number of siblings, free lunch eligibility, etc.; Academic performance data such as weekly studying hours, exam scores on reading and math \\
\hline
Outcome Column & Binary value indicating whether patient has died following hospitalization for vaccine reaction & Integer between 0 and 100 indicating score on writing exam \\
\bottomrule
\end{tabular}
\end{table*}

\begin{table*}[h]
    \centering
    \caption{Examples of LLM-generated Code Summaries}
    \label{tab:llm_code_summaries}
    \small
\begin{tabular}{ m{0.3\textwidth} m{0.6\textwidth}}
\toprule
\textbf{Data Operation}                    & \textbf{LLM-generated Summary} \\
\midrule
Dataset Loading                            & The code is loading data from the student\_exam\_scores.csv file. \\
\hline
Missing Value Imputation                   & The code is replacing any missing values in the "ParentEducation" column of the dataset with the most frequent value in that column. \\
\hline
Replacing Missing Values with Label        & This code is replacing any empty values in the 'EthnicGroup' column with the value 'unknown'.  \\
\hline
Outlier Removal                            & The data transformation is removing two rows (413 and 470) from the dataset that contain outlier values in the ReadingScore, WritingScore, and MathScore columns.  \\
\hline
One-hot Encoding                           & This transformation is creating new columns for each unique value of the "Gender" column in the dataset. For each row, the new columns will have a value of 0 or 1 depending on whether the row matches the value of the original column. \\
\hline
Transformation from Categorical to Numeric & The values in the "TestPrepCourse" column of the dataset are being replaced with 0 or 1 depending on whether the value is "none" or "completed".  \\
\hline
Feature Filtering                          & This code is removing the column named "PracticeSport" from the dataset. \\
\hline
Dataset Splitting                          & The data from the dataset is being split into two parts: X\_train and X\_test. X\_train contains the data from the dataset, excluding the column "WritingScore". The data in X\_train is then used to train a model. X\_test contains the same data as X\_train, but also includes the column "WritingScore". The data in X\_test is then used to test the model. \\
\bottomrule
\end{tabular}
\end{table*}